\documentclass{article}

\usepackage{arxiv}

\usepackage[utf8]{inputenc} 
\usepackage[T1]{fontenc}    
\usepackage{hyperref}       
\usepackage{url}            
\usepackage{booktabs}       
\usepackage{amsmath}
\usepackage{amsfonts}       
\usepackage{nicefrac}       
\usepackage{microtype}      
\usepackage{lipsum}
\usepackage{fancyhdr}       
\usepackage{graphicx}       
\graphicspath{{media/}}     

\title{Abnormal CO$_2$ and H$_2$O Diffusion in CALF-20(Zn) Metal-Organic Framework Angstropores
}

\author{
  Yann Magnin\\
  TotalEnergies, OneTech, R\&D, CSTJF, Pau, 64018, France.\\
  \texttt{\{Yann Magnin\}yann.magnin@totalenergies.com} \\
  \And
  Estelle Dirand \\
  TotalEnergies, OneTech, Power R\&D, Le Playground, Paris Saclay, 91400, France.\\
  \AND
  Guillaume Maurin\\
  ICGM, Univ. Montpellier, CNRS, ENSCM, Montpellier, 34293, France.\\
  \And
  Philip L. Llewellyn\\
  TotalEnergies, OneTech, Sustainability R\&D, CSTJF, Pau, 64018, France.\\
}

\begin{document}
\maketitle

\begin{abstract}
Carbon mitigation is one challenging issue that the world is facing. To tackle deleterious impacts of CO$_2$, processes emerged, including chemisorption from amine based solvents, and more recently physisorption in porous solids. While CO$_2$ capture from amine is more mature, this process is corrosive and detrimental for environment. Physisorption in Metal-Organic Frameworks (MOFs) is currently attracting a considerable attention, however the selection of the optimum sorbent is still challenging. While CO$_2$ adsorption by MOFs have been widely explored from a thermodynamics standpoint, dynamical aspects remain less explored. CALF-20(Zn) MOF was recently proposed as a promising alternative to the commercially used CO$_2$ 13X zeolite sorbents, however, in-depth understanding of microscopic mechanisms originating its good performance still have to be achieved. In this report, we deliver a microscopic insight of CO$_2$ and H$_2$O in CALF-20 by atomistic simulations. CALF-20(Zn) revealed to exhibit unconventional guest-host behaviors that give rise to abnormal thermodynamic and diffusion. The hydrophobic nature of the solid leads to a low water adsorption enthalpy at low loading followed by a gradual increase, driven by strong water hydrogen bonds, found to arrange as quasi 1D water wires in MOF porosity, recalling water behavior in carbon nanotubes and aquaporins. While no super-diffusion found, this behavior was shown to impact diffusion along with guests loading, with a minimum correlated with inflection point of adsorption isotherm corresponding to wires formation. Interestingly, diffusion of both CO$_2$ and H$_2$O were also found to be of the same order of magnitude with similar non-linear behaviors.
\end{abstract}


\bigskip

Fossil carbon avoidance, greenhouse gas compensation, carbon capture, \textit{etc}, are important puzzle pieces in the current carbon reduction strategy \cite{regufe2021}. However, carbon capture processes \cite{siegelman2021} represent an expensive brick in carbon management chains such as CCS (carbon capture and storage) or CCU (carbon capture and utilization) still limiting their massive deployment worldwide. Significant research and development is thus devoted to finding economic alternatives to the canonical amine scrubbing technique, where the thermal regeneration step tends to be highly energy consuming. Physisorption technique is currently of interest, although they equally suffer from their own disadvantages such as insufficient CO$_2$ product purity. To overcome such a limitation, research relates to intensification by developing rapid capture processes. Indeed, physisorption can be accelerated with minute-long cycles in both rapid-PSA (pressure swing adsorption) and rapid-TSA (temperature swing adsorption) \cite{subramanian2019,subraveti2019}. In these techniques, it is key to gain insight into the CO$_2$ adsorption microscopic mechanisms, not only from a thermodynamics point of view, but also from a diffusion standpoint, more rarely considered \cite{sharp2021}, although this is an important parameter in CO$_2$ capture from flue gas and from air. In addition, these carbon capture processes most often operate in presence of humidity and there is also a critical need to understand the water adsorption and diffusion in the selected sorbents \cite{yazaydin2009,hu2014,kolle2021,magnin2022,veldhuizen2023}. More generally, many adsorption processes, especially rapid capture technologies require efficient molecular transport in order to reach sorbent equilibrium in a limited time \cite{ruthven2012}. Up to now, diffusion resistances through scales being in nano-, meso- and macropores \cite{rouquerol1994,kondrashova2017} (in addition to the role of sorbent surfaces \cite{ruthven2012,remi2016}) are still elusive and diffusion phenomena through pore scales still need to be more intensively explored \cite{phan2020,magnin2020,berthonneau2021}. While continuum approach can be used for modeling convective fluid flow in macropores \cite{jarvis1991,regaieg2023}, diffusion in angstropores (\textit{i.e} sub-nanopores with sizes of about few molecular diameters) is a nascent field of research, \cite{beerdsen2006,bocquet2020} presenting unexpected and counter-intuitive physico-chemical behaviors \cite{falk2015,algara2015,bousige2021,kavokine2022,kapil2022}. Herein, we explored in-depth the diffusion of both CO$_2$ and H$_2$O in CALF-20(Zn), a prototypical MOF recently reported by Shimizu et al. \cite{lin2021}. This MOF structure is made of 1,2,4-triazolate-bridged zinc(II) ions layers pillared by oxalate ions to form a three-dimensional lattice encompassing a pore size of 0.6-0.7 nm, SI Appendix, Fig. \ref{fig:figSI1}. Thanks to its stability to humidity, its ability to capture CO$_2$ from wet flue gases (as well as other harsher contaminants NO$_x$, SO$_x$), and its rather simple and cheap synthesis \cite{lin2021}, CALF-20 has been demonstrated as a solid alternative sorbent to the currently used zeolite 13X \cite{chen2014}, even in the presence of relative humidity below 40\% to preferential adsorbtion. From a process perspective, this observation paves the way towards the implementation of CALF-20 in a moisture swing or humidity swing technology, as already proposed by the company Svante in their rapid TSA process \cite{siegelman2021,dhoke2021}. However, so far there is still a lack of understanding of the mechanisms at the origin of this excellent level of performance. Therefore this calls for an in-depth atomistic exploration of the thermodynamics and kinetics of both CO$_2$ and H$_2$O in CALF-20. To address this objective, we deployed a combination of force field-based grand canonical Monte Carlo (GCMC) and molecular dynamics (MD) approaches that revealed rather unusual adsorption and diffusion mechanisms for both guests. Typically, the first CO$_2$ molecules were found to be adsorbed in the center of the pores rather than at specific sites close to pore walls or metal centers \cite{lin2021}, while water was shown to form a quasi 1D molecular wires rather than clusters, recalling the water arrangement in single wall carbon nanotubes with diameter bellow 1 nm \cite{joseph2008,dalla2016,hummer2001}. Further we reveal that both the self and corrected diffusion coefficients are of same order of magnitude for the two guests molecules, and interestingly their concentration dependence follows an unconventional non-linear trend. A better understanding of such unconventional properties may allow to improve process modeling, formulate next polymorphs design and target material improvements and cost reductions for challenging sorption-based separations.\\

\begin{figure}
\centering
\includegraphics[width=0.9\linewidth]{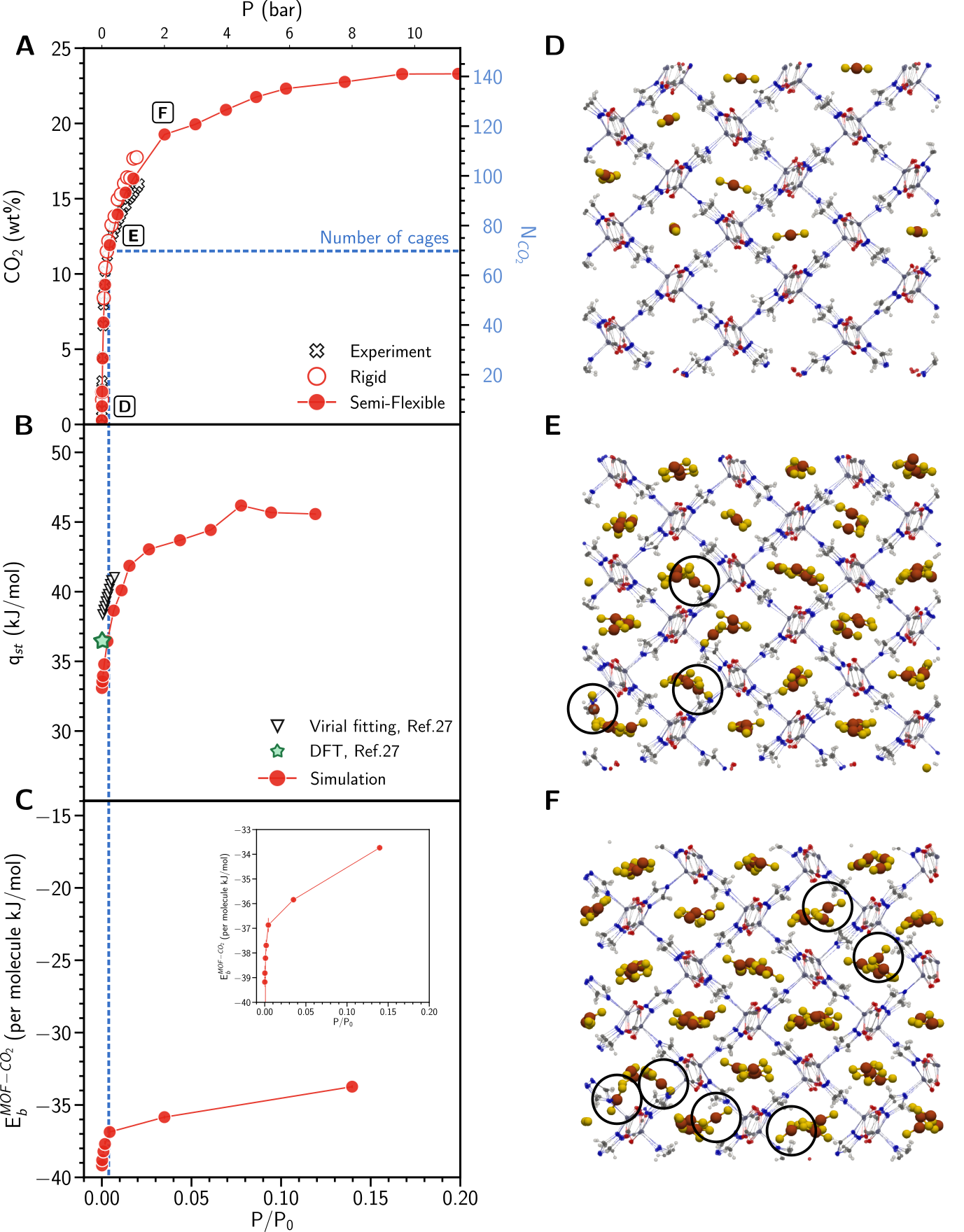}
\caption{A. CO$_2$ adsorption isotherm at $T$ = 293.15K. Red circles correspond to GCMC simulations, open cross symbols to experiment \cite{lin2021}. The horizontal dashed blue line corresponds to the total number of cages in the MOF atomistic structure, the vertical one to the relative pressure at which all cages are filled by one CO$_2$ molecule. B. CO$_2$ isosteric enthalpy of adsorption at $T$ = 293.15K. The green star corresponds to density functional theory reported in \cite{lin2021} and the open triangles to enthalpy of adsorption from virial fitting method \cite{lin2021}. C. Per molecule MOF-CO$_2$ biding energy. The inset corresponds to a zoom of the figure. D-F Illustration of the preferential arrangements of CO$_2$ molecules simulated by GCMC in CALF-20 at different $P/P_0$=5$\times$10$^{-5}$ (D), 4$\times$10$^{-3}$ (E), 3.5$\times$10$^{-2}$ (F). Black circles correspond to cages occupied by more than one CO$_2$ molecule.}
\label{fig:fig1}
\end{figure}

\noindent The GCMC CO$_2$ adsorption isotherms, calculated using both rigid (no MOF atoms vibrations, orange open circles) and flexible (MOF atoms vibrating, orange circles) CALF-20 framework at $T$=293.15 K are compared to the corresponding experimental data reported in \cite{lin2021} (open cross symbols) in Fig. \ref{fig:fig1}\textit{A}. The simulated isotherms are of Langmuir type, characteristic of nanoporous adsorbents, with no effect in MOF being considered as a rigid or flexible framework. In Fig. \ref{fig:fig1}\textit{B}, we show the isosteric enthalpy of adsorption, found to increase as a function of the partial pressure $P/P_0$. This energetic parameter was determined from the fluctuation method, 
\begin{equation}
q_{st}\ =\ k_BT\ - \frac{\partial <U>}{\partial <N>}\ =\ k_BT\ - \frac{<UN>-<U><N>}{<N^2>-<N>^2},
\end{equation}
with $q_{st}$ the isosteric enthalpy of adsorption, $k_B$ the Boltzmann constant, $N$ and $U$ the guests number and potential energy, respectively. At highly dilute conditions, the simulated $q_{st}\ \sim$ 35 kJ/mol, in agreement with the literature (green stars as well as open triangles) \cite{lin2021}, followed by an increasing profile reaching $\sim$ 45 kJ/mol at large $P/P_0$. We have also determined the guest-host biding energy, after annealing the system. It is defined such as $E^{G-H}_b\ =\ E^{G+H}-E^{H}-E^{G}$, where $E^{G+H}$ corresponds to the potential energy of the host (H) filled by guest adsorbates (G), $E^{H}$ is the host energy after removing guest molecules, and $E^{G}$ is the energy of guest molecules after removing the host. $E^{G-H}_b$ thus corresponds to the binding energy of the fluid medium interacting with the pore walls. Below $P/P_0 \ \sim$ 3.10$^{-3}$, $E^{G-H}_b$ is shown to strongly increase (less attractive MOF surfaces), Fig. \ref{fig:fig1}\textit{C}, while $q_{st}$ continuously increases. This behavior is attributed to G-G interactions driving the increase of the overall enthalpy of adsorption. In the CALF-20, CO$_2$ molecules are found to arrange in the near center of MOF cages \cite{lin2021}, Fig.\ref{fig:fig1}\textit{D}. When $P/P_0 \ \sim$ 3.10$^{-3}$, almost all MOF cages are occupied by one CO$_2$ molecule, Fig.\ref{fig:fig1}\textit{E}. This is further confirmed by the number of adsorbates found to match the total number of cages, dashed blue line in the Fig.\ref{fig:fig1}\textit{A}. At larger pressure, the overall $E^{G-H}_b$ decreases (less negative energy), corresponding to configurations with additional guest molecules in some cages, Fig.\ref{fig:fig1}\textit{F}. The later arrangement will be more discussed below in the light of the radial distribution function.\\

\begin{figure}
\centering
\includegraphics[width=0.9\linewidth]{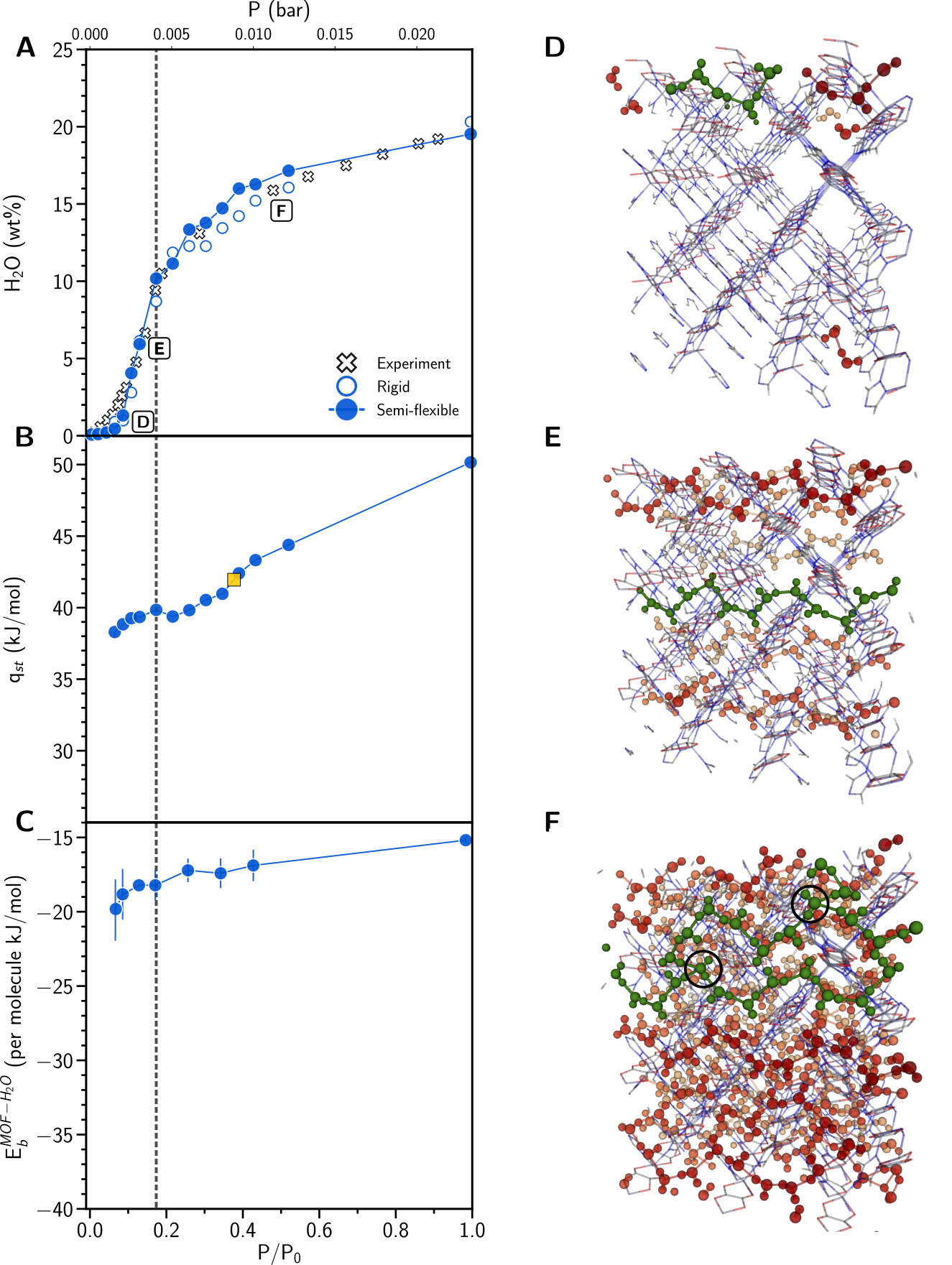}
\caption{A. Water adsorption isotherm at $T$ = 293.15K. Blue circles correspond to simulations, open cross symbols to experiments \cite{lin2021}. B. Water isosteric enthalpy of adsorption at $T$ = 293.15K. The yellow square corresponds to the enthalpy of adsorption for CO$_2$ at $P_{CO_2}$=1 bar C. Per molecule MOF-H$_2$O biding energy. D-F Illustration of the preferential arrangements of water molecules simulated by GCMC calculations in CALF-20 at different $P/P_0$. D. $P/P_0$ = 0.05, water molecules are found to be isolated and/or forming multimers through hydrogen bonds. For a sack of visibility, each molecules or multimers were colored depending on their positions on $x$ axis. E. $P/P_0$ = \textbf{0.15}, water molecules start arranging as wires crossing MOF cages. Each water molecule present 2 H bonds, leaving one H dangling bond. F. $P/P_0$ = 0.5, wires branching throughout the MOF porosity. The green chain is made of several wires interconnected by water molecules with tree hydrogen bonds (black circles).}
\label{fig:fig2}
\end{figure}

\noindent The same analysis was then conducted for water adsorption at same $T$. The H$_2$O isotherm presents a S-shape, Fig. \ref{fig:fig2}\textit{A}. As for CO$_2$, the consideration of a rigid (open blue circles) and flexible (blue circles) MOF framework does not lead to differences in the calculated isotherms. At low loading, $q_{st} \sim$ 40 kJ/mol, a value bellow the water vaporization enthalpy, denoting an hydrophobic nature of the structure, Fig. \ref{fig:fig2}\textit{B} (further confirmed by the S-shape isotherm). The following increase is thus attributed to G-G interactions, confirmed by the G-H biding energy presenting a smooth increase (less attractive pore surfaces), Fig. \ref{fig:fig2}\textit{C}. At highly dilute conditions, water is shown to adsorb pore sites, mostly as isolated molecules and/or small multimers, Fig. \ref{fig:fig2}\textit{D}. When the loading increases, formation of hydrogen bonds drives the increase in $q_{st}$ (Fig. \ref{fig:fig2}\textit{B}). However, surprisingly water molecules are not found to form clusters (as usually reported for many MOF structures \cite{ghosh2004,ding2019,hanikel2020,cong2021,magnin2022}), but arrange as a quasi 1D molecular wires propagating through MOF cages, Fig. \ref{fig:fig2}\textit{E}. This uncommon molecular arrangement recalls the behavior of water in small diameter single wall carbon nanotubes (SWCNT), where 1D water wires were shown to result from strong confinement in highly hydrophobic solid pores, where water/water interactions dominate \cite{striolo2005}. As for water adsorbing in CALF-20 angstropores, H$_2$O adsorption in SWCNT shows an increasing $q_{st}$ along with $P/P_0$, attributed to the formation of a strong water H-bonds network \cite{joseph2008,dalla2016,hummer2001}. Such a behavior, here shown in 3D pores of CALF-20 may arise (as for 1D SWCNT) from the relative hydrophobic nature of the MOF, in addition to the high confinement of the MOF cages. Note that the increase of the wires number as well as their lengths along with $P/P_0$, may contribute to smoothly decrease $E^{G-H}$ (less negative) due to possible uncommensurability of water H-bonds in wires with pore surface atoms of the MOF. At large $P/P_0$, percolation occurs, with neighbor wires found to interconnect, Fig. \ref{fig:fig2}\textit{F}. This behavior is more detailed by the cluster analysis discussed bellow from MD calculations. It is worth noting that $q_{st}$ for the two guests species (CO$_2$ and H$_2$O) present similar values (Fig. \ref{fig:fig1}\textit{B} and Fig. \ref{fig:fig2}\textit{B}) despite the notable difference in their respective G-H binding energies (Fig. \ref{fig:fig1}\textit{C} and Fig. \ref{fig:fig2}\textit{C}) attributed to G-G interactions. We can also notice that $q^{CO_2}_{st}\sim q^{H_2O}_{st}$ at $P_{CO_2}\sim$ 1 bar at a relative humidity found at about 42\% (yellow square in Fig. \ref{fig:fig1}\textit{B}), in agreement with the result reported in \cite{lin2021} at 40\%.\\

\begin{figure}
\centering
\includegraphics[width=1.0\linewidth]{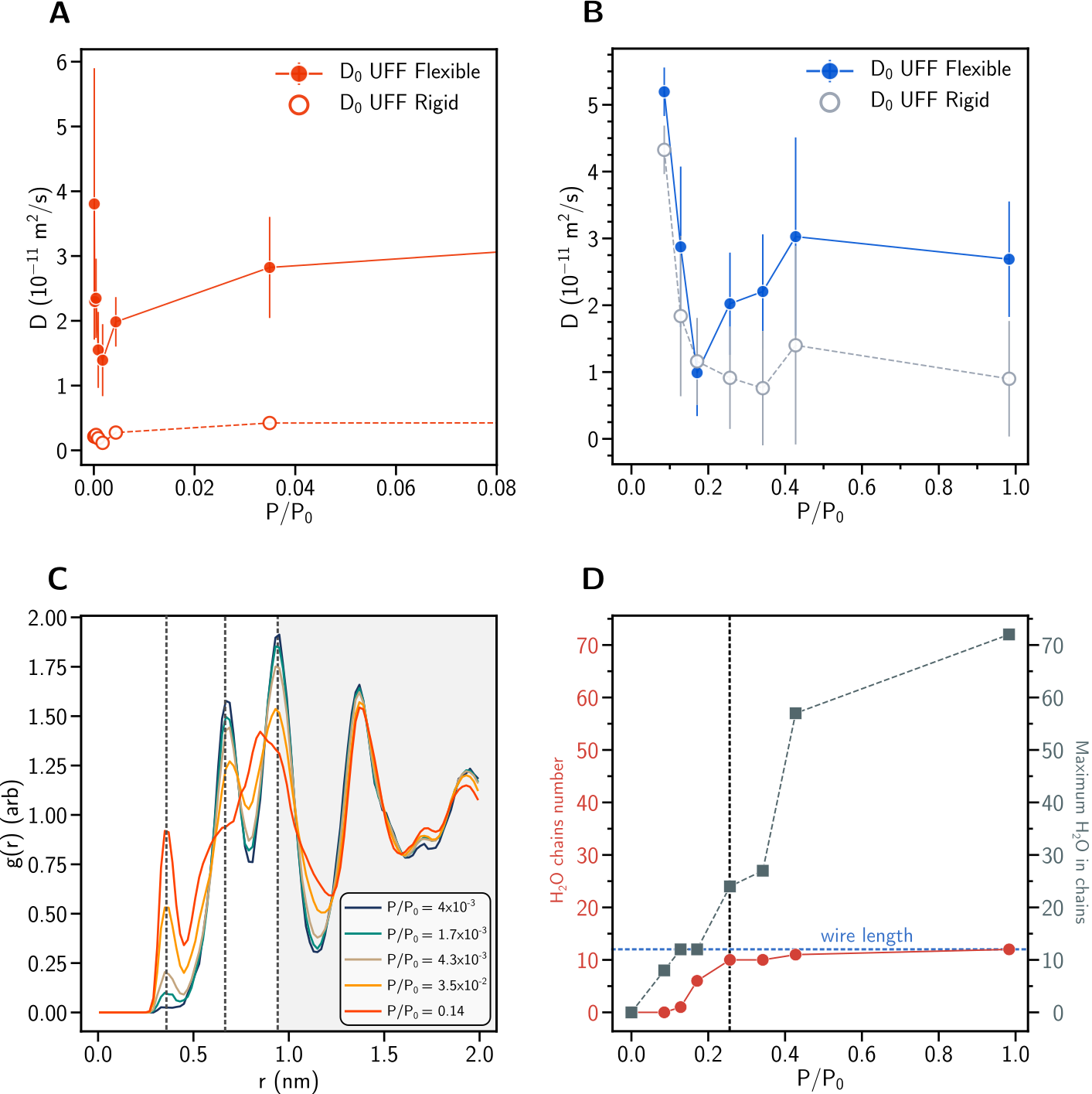}
\caption{A. Simulated corrected diffusion coefficient for CO$_2$ as a function of $P/P_0$. B. Simulated corrected diffusion coefficient for H$_2$O as a function of $P/P_0$. C. Radial distribution function of the CO$_2$ center of mass for different partial pressures. The gray background corresponds to the second neighbor area. D. Water wires analysis, with orange circles corresponding to the number of wires into the structure. The gray squares correspond to the mean number of water molecules per wire, surrounded by the maximum (upper dashed line) and minimum (lower dashed line) number of water molecules spanning all wires in the structure. The gray area denotes the gap between the smaller and the larger wire in structures.}
\label{fig:fig3}
\end{figure}

\begin{figure*}
\centering
  \includegraphics[height=6cm]{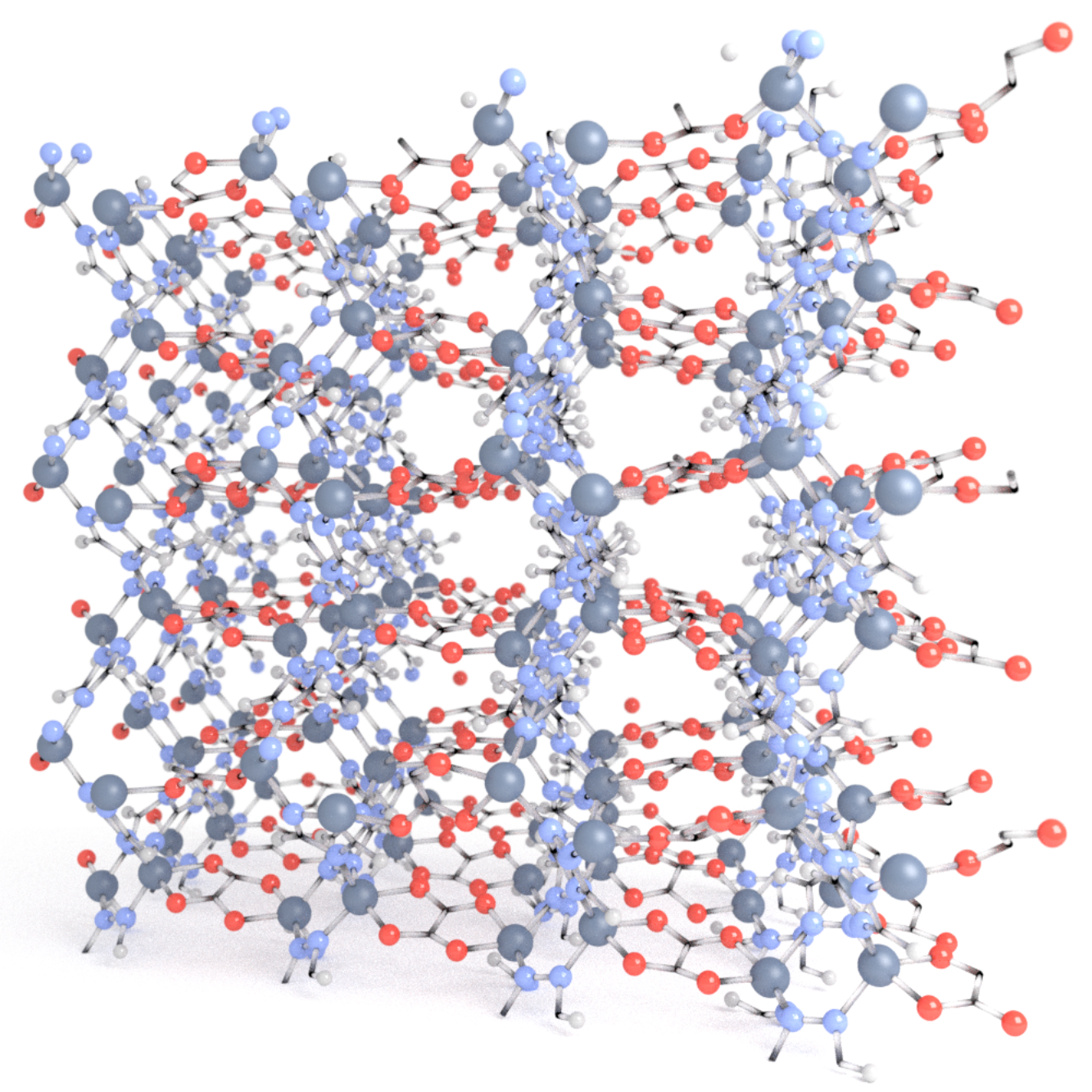}
  \caption{Atomistic structure of CALF-20 made of 4$\times$3$\times$3 super-cells. The CALF-20(Zn) structure is made of zinc oxide nodes, bridged by triazole and oxalate linkers with a resulting sub-nanoporous structure formed by cages with diameters of $\sim$0.6 - 0.7 nm and a surface area of about $\sim$442 m$^2$/g \cite{wei2022}.}
  \label{fig:figSI1}
\end{figure*}

\begin{figure*}
\begin{center}
\includegraphics[scale=0.4]{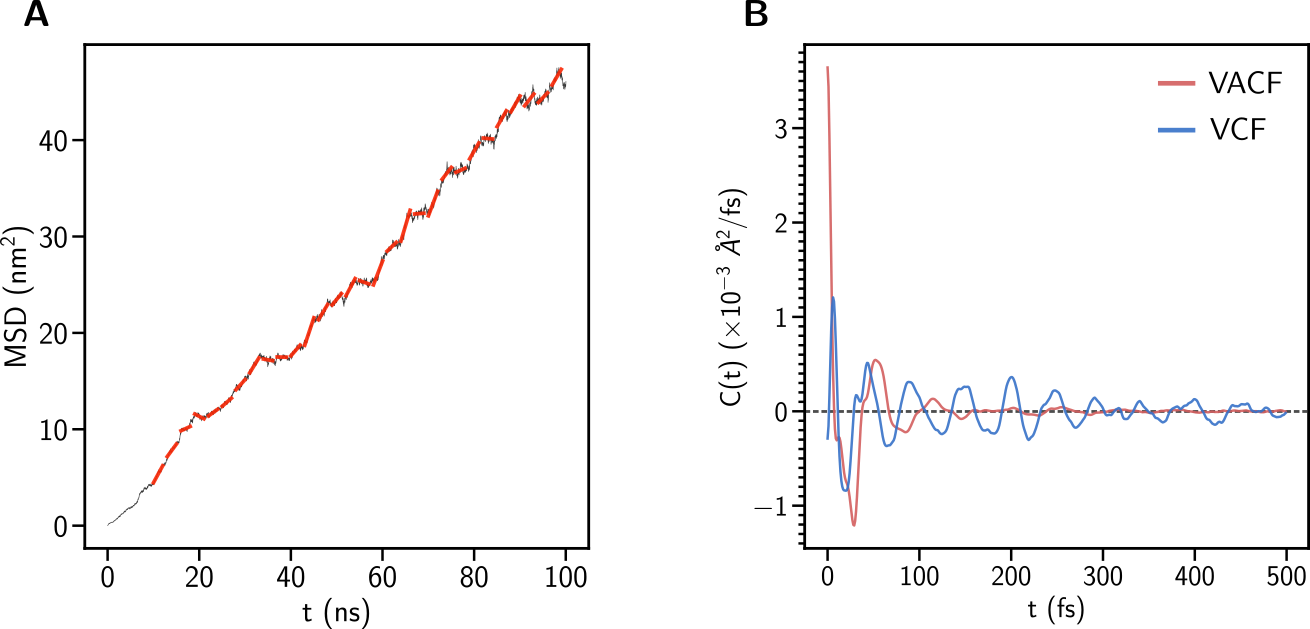}
\end{center}
\caption{\label{fig:figSI3} A. Mean square displacement of CO$_2$ molecules at $T$=293.15K and $P$=1 bar (black line) calculated from equilibrium MD. The red segments corresponds to fits of the MSD plot on short time scale. B.Velocity auto-correlations and velocity correlation of CO$_2$ molecules at $T$=293.15K and $P$=1 bar (red and blue lines, respectively).}
\end{figure*}

\begin{figure}[ht]
\begin{center}
\includegraphics[height=6.5cm]{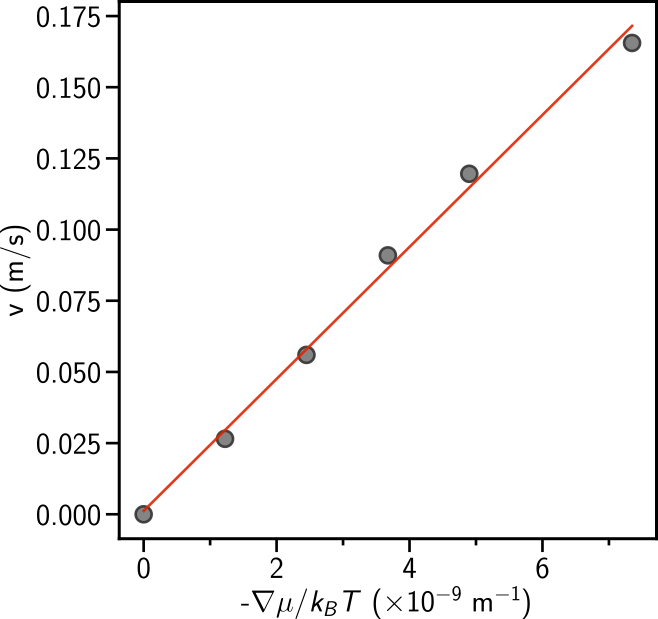}
\end{center}
\caption{\label{fig:figSI4} Velocity plotted as a function of different driving forces by out of equilibrium molecular dynamics (brown dots). The red line corresponds to the fit with a slope corresponding to $D_0$.}
\end{figure}

\begin{figure}
\centering
\includegraphics[height=6.5cm]{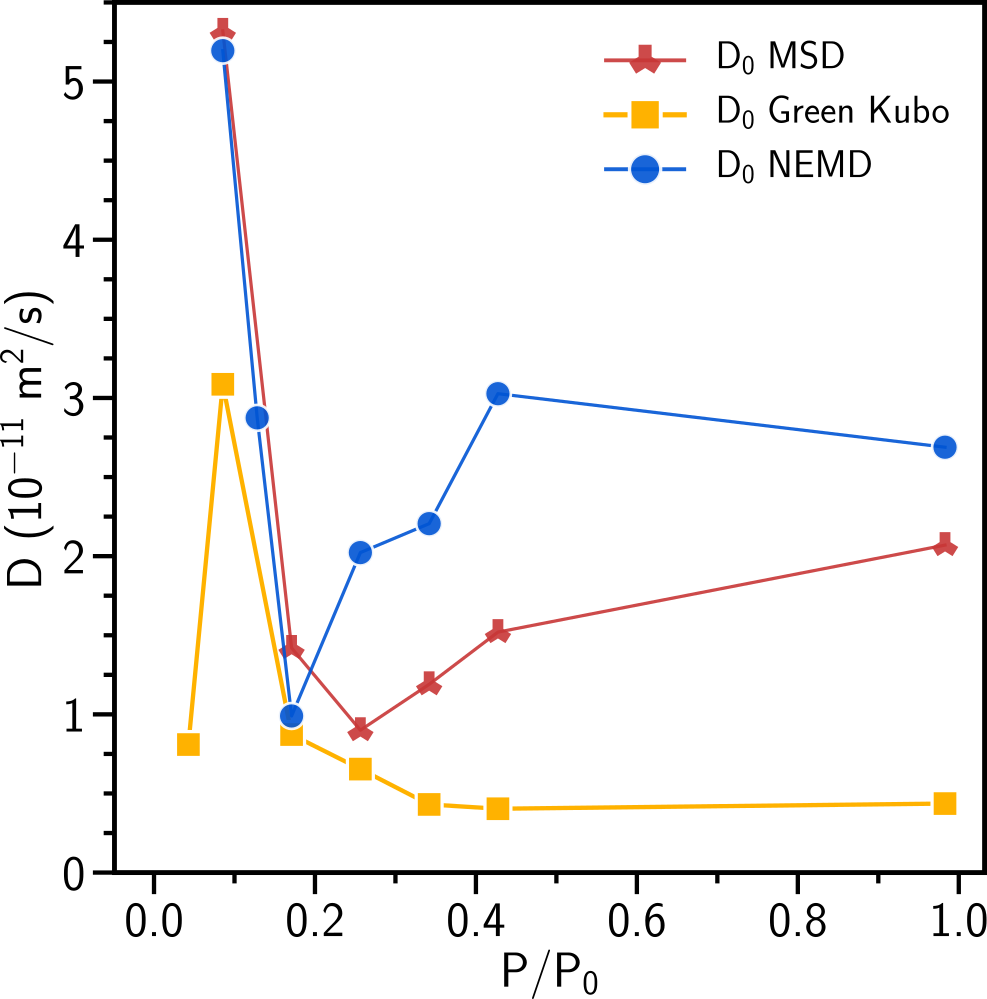}
\caption{Corrected diffusion coefficient of water molecules in flexible CALF-20 by UFF force field using different methods. The red crosses corresponds to $D_0$ from the MSD method, the yellow squares to the Green-Kubo approach and the blue circles to the NEMD technique. For a sack of clarity we remove error bars.}
  \label{fig:figSI5}
\end{figure}

\noindent While CO$_2$ and H$_2$O thermodynamics was shown above to present some behaviors differing from several other MOFs, such differences also impact guests mobility, found to present abnormal non-linear features in diffusion coefficients, with minimum in diffusion at intermediate guests loading. To insight it, MD simulations were further performed to determine the self and corrected diffusion coefficients, labeled such as $D_s$ and $D_0$, respectively. To do so, we used the mean square displacement, the Green-Kubo relation and a non-equilibrium molecular dynamics method (NEMD), based on transport theory \cite{falk2015}. The different methodologies used and computational details are described in the SI Appendix, Fig. \ref{fig:figSI3} \textit{A,B} and \ref{fig:figSI4}. A comparison of the application of the different approaches is presented in SI Appendix, Fig. \ref{fig:figSI5}, showing consistency in diffusion coefficients calculated from the different methods.\\
In order to check the impact of the MOF flexibility on CO$_2$ diffusion, we then performed calculations in both rigid and flexible frameworks, Fig. \ref{fig:fig3}\textit{A}. We thus show that accounting for thermal effects in the solid tends to boost the CO$_2$ diffusion (orange circles) by one order of magnitude compared to the rigid approximation (open orange circles). Note that such a behavior has already been reported for several other porous solids \cite{yang2022,terranova2016}. Self and corrected diffusion coefficients were then compared along with $P/P_0$ for the two guest species from MSD and NEMD techniques (SI Appendix, SI Text section A). The two diffusion coefficients were found to present the same order of magnitude. Based on the Green Kubo relation, we can express the corrected diffusion coefficient given by,

\begin{equation}
\label{eq:GK}
D_0 = \frac{1}{d N}\  \left\lbrace\ \sum_{i=1}^{N}\ \int_{0}^{\infty} \left\langle \textbf{v}_i(t).\textbf{v}_i(0) \right\rangle dt\ +\ \sum_{i\neq j}^{N}\int_{0}^{\infty} \left\langle \textbf{v}_j(t).\textbf{v}_i(0) \right\rangle dt\right\rbrace.
\end{equation}

\noindent In it, $d$ is the dimension of the system, $N$ the number of guests, $\textbf{v}$ the velocities of molecules $i$ and $j$, and angular brackets denote the ensemble average of the velocity correlation. The first summation in \eqref{eq:GK} corresponds to $D_s$, and the second to the cross correlations. For CO$_2$ diffusion, the cross term is found to be negligible and $D^{CO_2}_s \sim D^{CO_2}_0$, SI Appendix, S5 \textit{A}. This behavior originates from the strong guest confinement in CALF-20 angstropores, where the G-H interactions dominate as compared to G-G interactions. Doing the same for water molecules, $D^{H_2O}_s\ \sim D^{H_2O}_0$ up to $P/P_0\ \sim 0.2$ (inflection point in the isotherm), SI Appendix, S5 \textit{B}. While water behaves as independent molecules, G-H represent the main interactions (despite a relatively low biding energy), and the limited number of G neighbors not induces large cross correlation effect in equation \eqref{eq:GK}. However, once water wires form ($P/P_0 \sim$ 0.2), collective interactions can no longer be neglected, and a deviation can be observed ($D^{H_2O}_s\ \neq D^{H_2O}_0$), SI Appendix, S5 \textit{B}. MOF semi-flexibility versus rigid MOF has also been compared for water, showing a moderate effect. A deviation is however observed at $P/P_0\ \sim 0.2$, suggesting that solid vibrations promote water diffusion when wires are formed, Fig. \ref{fig:fig3}\textit{B}.\\

\noindent $D^{CO_2}_0$ from NEMD (in flexible MOF) shows a non-monotonic behavior, with a minimum diffusion at $P/P_0\ \sim$ 10$^{-3}$. When $P/P_0\ <$ 10$^{-3}$, a few isolated molecules have been shown to interact with low energy sites of the solid. Thanks to the available free pore volume, molecules have a probability to diffuse, hopping from one cage to another. When $P/P_0\ \sim$ 10$^{-3}$, almost all cages are occupied by one guest molecule (dashed blue line in Fig. \ref{fig:fig1}\textit{A}). In such a configuration, CO$_2$ molecules are constrained to share a cage with a neighbor CO$_2$. This situation is thermodynamically not favorable and originates a decrease of the guest mobility, reflected by the minimum in $D^{CO_2}_0$, Fig. \ref{fig:fig3}\textit{A}. This is confirmed by the slope change shown in $E^{G-H}$ (vertical dashed blue line in Fig. \ref{fig:fig1}\textit{C}) and by the CO$_2$ center of mass radial distribution function (RDF), Fig. \ref{fig:fig3}\textit{C}. In the RDF, the second and third peaks correspond to CO$_2$ molecules in first neighbor cages on $z$ and ($x,y$) directions respectively, and the gray area corresponds to second neighbor cages. However, the first peak below 0.5 nm corresponds to additional guest molecules in some cages, already hosting one CO$_2$ molecule, further reflected by the increase of the first peak in the RDF along with the partial pressure increase. Hence, just above $P/P_0\ \sim$ 10$^{-3}$, some cages are occupied by more than one CO$_2$ molecule, sharing weaker interactions with solid surfaces, increasing the CO$_2$ diffusion coefficient. At larger $P/P_0$, the corrected diffusivity reach a constant value, typical of large pore volume occupation, where diffusion is limited by guest collisions.

\noindent Whilst water diffusion also presents a minimum at $P/P_0\ \sim$ 0.2, mechanisms behind diffusion behaviors are different from those depicted for CO$_2$. In the Fig. \ref{fig:fig3}\textit{D}, we can observe the evolution of wires number (orange circles), as well as their constitutive maximum number of molecules (gray squares). In the atomistic MOF structure used for simulations, a water wire crossing the structure throughout a channel (made of connected MOF cages along one direction) counts 12 water molecules. Thus, water does not present wires below $P/P_0 \sim$ 0.2. In this situation, the growth of multimers ($<$ 12 water molecules, horizontal dashed line) is accompanied by a decrease in the water mobility, with a minimum in diffusion at around $P/P_0 \sim$ 0.2 (Fig. \ref{fig:fig3}\textit{B}). At such a pressure, 10 wires are found to occupy the 12 MOF channels, with a wire number reaching a maximum in Fig. \ref{fig:fig3}\textit{D}. At slightly higher pressure (vertical dashed line), additional water molecules are found with extra-hydrogen bonds bridging some molecular wires together (Fig. \ref{fig:fig2}\textit{E}), corresponding to the early stage of the water percolation. These bridging molecules are less stable than ones within a wire, increasing $D^{H_2O}_0$. At $P/P_0\ \sim$ 0.4, wires are largely inter-connected through cages (Fig. \ref{fig:fig2}\textit{F}) and water medium is clearly percolated driving the saturation in diffusion. Despite similarities with confined water in SWCNT, H$_2$O diffusion in CALF-20 differs with diffusion coefficient found to be far lower than that of the bulk. This may be attributed to the pore network presenting tortuosity compared to straight nanotubes, making water wires not fully 1D. In addition to the fact that SWCNT are more hydrophobic than CALF-20 with enthalpy of adsorption ($q^{SWCNT}_{st}\sim$10 kJ/mol \cite{striolo2005} and $q^{CALF-20}_{st}\sim$40 kJ/mol) and that water percolation in SWCNT cannot occur as compared to the CALF-20 situation.\\ 

\noindent Comparing the two guests we can notice that $D^{CO_2}_0$ and $D^{H_2O}_0$ present the same order of magnitude, differing from number of other MOFs where water usually shows a smaller mobility compared to CO$_2$. This behavior relates to specific properties of the CALF-20, where both CO$_2$ and H$_2$O show similar enthalpies of adsorption, presenting an increasing trend with gas pressure related to G-G interactions. Despite close enthalpies of adsorption, a simple comparison of such a parameter for both guests molecules shows that CO$_2$ adsorption is favored over H$_2$O bellow $P/P_0\ \sim$ 0.42 (in agreement with already reported results \cite{lin2021}). At low humidity water was found as isolated molecules and/or small multimers ($P/P_0\ <$ 0.2), and unconnected wires ($P/P_0\ \sim$ 0.4) found to weakly interact with MOF pore surfaces. In such conditions, CO$_2$ presenting a smaller enthalpy of adsorption may displace water from CALF-20 angstropores. However, when  the relative humidity become larger than $\sim$ 0.4, water medium was found as a percolated network. In such an arrangement, water may be blocked through MOF cages, and CO$_2$ at 1 bar is not able to break water H bonds and displace the water medium. A corresponding situation may explain why water adsorption is favored compared to CO$_2$. Beyond such a mechanistic analysis, CALF-20(Zn) is currently receiving a particular attention due to its unconventional physico-chemical properties, in addition to its simple synthesis. Although this work focuses on the thermodynamics and kinetics of two guests molecules in CALF-20 intra-pores, we believe that such atomistic bricks could help in taking advantages of such unconventional behaviors for future applications.

\noindent A supercell model of CALF-20 made of 4$\times$3$\times$3 unit-cells (SI Appendix, Fig. \ref{fig:figSI1}) was considered for all GCMC and MD simulations. The MOF framework was treated as flexible (\textit{i.e} each atom of the MOF is free to move under thermal excitations, while the MOF volume is kept constant) for both thermodynamics and kinetics studies, using the universal force field UFF for the intramolecular bonding, bending, and dihedral terms to describe the MOF framework. In this later case, this is even more important since it is well-documented that the consideration of the flexibility of the MOF framework is of utmost importance to describe guests diffusion \cite{yang2022,salles2011,wang2021,obliger2019}. For comparison the adsorption isotherms and diffusion coefficients were also calculated with a rigid framework considering the structure reported in \cite{lin2021}. The CO$_2$ and H$_2$O adsorption isotherms and enthalpies of adsorption were determined by GCMC with the lammps code \cite{plimpton1995}. For each calculation, deviation from the ideal gas was corrected from the Soave-Redlich-Kwong model \cite{peng1976}. Diffusion properties were assessed by MD simulations in the canonical ensemble. Consistency of diffusion coefficients obtained was verified by calculations made in both equilibrium and out of equilibrium conditions (SI Appendix, SI Text, section A) \cite{frentrup2012}, with lammps. All atoms of the MOF framework were treated as charged Lennard Jones (LJ) interacting sites with LJ parameters taken from UFF \cite{rappe1992}. The REPEAT model \cite{campana2009} was used to assign point charges on each MOF atom as it was demonstrated to be well suited to reproduce both CO$_2$ and H$_2$O isotherms \cite{lin2021}. CO$_2$ molecules were modeled by the long range Calero's pair potential \cite{calero2009}, shown to reproduce well the experimental isotherms at ambient conditions using the CALF-20 structure provided in \cite{lin2021}, while water was modeled from the four sites TIP4P-Ew \cite{horn2004} potential. A 1 nm cut-off was applied for all LJ interactions, in addition, a long-range Coulombic solver (Ewald summation) was considered. Each interatomic LJ parameters was determined from the Lorentz Berthelot mixing rule. Simulation details are provided in the SI Appendix, SI Text, section B.\\

\clearpage
\Large
\noindent \textbf{Supporting Information}\\
\normalsize

\noindent \textbf{Diffusion theory}\\
In equilibrium bulk phase, thermal energy induces particle diffusion through Brownian's motion. For a single tagged particle, wa can assess the self-diffusion coefficient $D_s$, depending on the fluid temperature and density \cite{falk2015}. Self-diffusion is accessible from different experimental techniques like NMR, QENS or IFM, and from the plot of the time dependence of the mean square displacement (MSD) simulated for a tagged particle \cite{rouquerol2013,kaerger2015}. MSD  is based on Einstein's relationship, stating that the diffusion coefficient is given by the mean-square displacement of molecules along time such as,

\begin{eqnarray}
\label{eq:msd1}
D_s &=& \frac{1}{2d N}\ \lim_{t\rightarrow \infty} \frac{d}{dt} \left\langle \sum_{i=1}^{N} (\textbf{r}_i(t)-\textbf{r}_i(0))^2 \right\rangle.
\end{eqnarray}

In \eqref{eq:msd1}, $d$ is the dimension of the system, $N$ the number of particles, $r_i(t)$ the position of a particle $i$ at time $t$ and the angular brackets denote an ensemble average. For systems filled by molecules, the MSD increases linearly as a function of time due to molecules collisions and the average of the MSD slope allows to determine $D_s$, Fig. SI \ref{fig:figSI3}\textit{A}.\\ 
To account for collective interactions, corrected diffusion coefficient $D_0$ can be expressed by different methods. $D_0$ can be approached by MSD, while instead averaging each molecule positions, collective displacement is averaged from the fluid center of mass $R$. Thus, \eqref{eq:msd1} can be expressed in its corrected form as follow \cite{rouquerol2013},

\begin{eqnarray}
\label{eq:msdcoll}
D_0 &=& \frac{N}{2d}\ \lim_{t\rightarrow \infty} \frac{d}{dt} \left\langle (\textbf{R}(t)-\textbf{R}(0))^2 \right\rangle.
\end{eqnarray}

However, this approach presents strong fluctuations and required larger number of long simulations compared to $D_s$ \cite{falk2015}. Derivating \eqref{eq:msd1}, we can formulate the Green-Kubo equation (GK),

\begin{eqnarray}
\label{eq:msd2}
D_s &=& \frac{1}{d N}\  \sum_{i=1}^{N}\ \int_{0}^{\infty} \left\langle \textbf{v}_i(t).\textbf{v}_i(0) \right\rangle dt.
\end{eqnarray}

\noindent In \eqref{eq:msd2}, the dot product of velocities corresponds to the velocity auto-correlation function (VACF). In confined systems, the VACF presents an oscillating behavior, resulting from molecules collisions and molecules (back-)scattering due to their interactions with an adsorption site, Fig. \ref{fig:figSI3}\textit{B}.  

\noindent It stems from the above that molecules are also more or less correlated with the motion of their neighbors and particles diffusion is thus a collective property writing in GK formalism as a sum of $D_s$ \eqref{eq:msd2} (first sum in \eqref{eq:Dc}) and cross terms, integrating the velocity correlation function (VCF) (second sum in \eqref{eq:Dc}),

\begin{eqnarray}
\label{eq:Dc}
D_0 &=& \frac{1}{d N}\  \left\lbrace\ \sum_{i=1}^{N}\ \int_{0}^{\infty} \left\langle \textbf{v}_i(t).\textbf{v}_i(0) \right\rangle dt\ +\ \sum_{i\neq j}^{N}\int_{0}^{\infty} \left\langle \textbf{v}_j(t).\textbf{v}_i(0) \right\rangle dt\right\rbrace.
\end{eqnarray}

\noindent In highly confined environments (pore sizes of few molecular diameters), molecules mostly interact with solid pore surfaces, while guest interactions or their coordination remains more or less limited. In such conditions, molecules behave almost independently and the cross term in \eqref{eq:Dc} is negligible, $D_s\ \sim\ D_0$ cite{falk2015}. If such an approximation can be made for confined and dilute media, a special care has to be made with dense media, for example water, that tends to form clusters in (sub-)nanopores. The GK is challenging method to be converged. It depends on the integration time step and time length, where the infinite limit in integrals has to be approximated and large ensemble average is required.\\
Transport diffusion (non-equilibrium dynamics) \cite{frentrup2012} presents an interesting alternative to equilibrium methods described above. In such a technique, flux of molecules is driven by the application of an external driving force. In the steady regime, the flux of molecules can be expressed as follow,

\begin{eqnarray}
\label{eq:flux}
j\ =\ \rho \times v,
\end{eqnarray}

\noindent with $j$ denoting the molecules flux, $\rho$ the molecules density and $v$ their velocities. Noting that the driving force initiating the molecules flow can be noted such as,
 
\begin{eqnarray}
\label{eq:fonda1}
F\ =\ v\times f\ =\ - \nabla \mu,
\end{eqnarray}

\noindent where $v$ is the fluid velocity, $f$ the friction coefficient of fluid molecules and $F=-\nabla \mu$ the driving force, expressed as a gradient of chemical potential. Noting that $\mu=\mu_0+k_BT \ln(P/P_0)$ and coupling to \eqref{eq:fonda1}, we can express the first Fick's law,

\begin{eqnarray}
\label{eq:fonda2}
j &=& - \frac{k_BT}{f} \frac{d\ln(P)}{d\ln(\rho)} \frac{d\rho}{dr}\ =\  -D_t \nabla \rho,
\end{eqnarray}

\noindent with $D_t$, the transport coefficient. From \eqref{eq:fonda2},

\begin{eqnarray}
\label{eq:fonda3}
D_t &=& \frac{k_BT}{f} \frac{d\ln(P)}{d\ln(\rho)}\ =\ D_0\times \Gamma,
\end{eqnarray}

\noindent with $D_0$ the corrected diffusion coefficient and $\Gamma$ the thermodynamic factor (Darken's factor). For simulation purpose, \eqref{eq:fonda2} is reformulated with the driving force, $-\nabla \mu$. Noting that $D_0=k_BT/f$ and $\nabla \mu/k_BT=\nabla \ln(P)$, molecules velocity reads,

\begin{eqnarray}
\label{eq:osanger2}
v &=& -\frac{D_0}{k_BT} \nabla \mu.
\end{eqnarray}

\noindent To determine $D_0$ from NEMD simulations, we used \eqref{eq:osanger2}, applying a constant force $F=-\nabla \mu$ on the center of mass of guests molecules along one direction. During simulations, the center of mass velocity is average along time, and $D_0$ can then be fitted has shown in Fig. SI \ref{fig:figSI4}. It is important to note that the thermostat in molecular dynamics is turned off along the flow direction in order to not introduce a flow bias.

\textbf{Methods}
All simulations have been made in a triclinic CALF-20 atomistic structure of 4$\times$3$\times$3 super-cells with REPEAT point charges proposed by Shimizu et al. \cite{lin2021}, Fig. SI \ref{fig:figSI1}. Thermodynamics simulations (adsorption and isosteric enthalpy) has been calculated from Monte Carlo algorithm in the grand canonical ensemble (GCMC) \cite{frenkel2002}, in which the MOF structure randomly exchanges molecules with an ideal gas reservoir kept at constant guests chemical potential $\mu$, volume $V$ and temperature $T$ = 293.15K. A cycle consists in trying 500 molecules insertions or deletions, randomly chosen with 50\% probability, followed by 5 ps relaxation by molecular dynamics. GCMC simulations were run for $4\times 10^4$ Monte Carlo cycles for CO$_2$ and $1\times 10^5$ for H$_2$O adsorption. The first half cycles were used to equilibrate the system and remaining cycles were used for averaging guests density from equilibrium microstates.\\ 
GCMC simulations do not directly control the pressure, but the chemical potential, 

\begin{equation}
\mu\ =\ k_BT\ \ln\left(\frac{f\ \Lambda^3}{k_BT}\right),
\label{eq:muP}
\end{equation}

\noindent with $\Lambda$ the de Broglie thermal wave length, and $f$ the fugacity. The pressure is thus determined from $f$, whose form in Eq.\ref{eq:muP} depends on the ideal gas equation of state. In order to estimate the pressure, that may deviate from ideal gas under some thermodynamic conditions, fugacity has been corrected from the Soave-Redlich-Kwong (SRK) model \cite{peng1976}.\\
Inter-atomic parameters has been calculated from the Lorentz Berthelot mixing rule and can be found in lammps inputs shared with this paper (UFF and Dreiding available). The CO$_2$ and H$_2$O parameters can be found in the work of Calero et al. \cite{calero2009} and Horn et al. \cite{horn2004}, respectively. Force field cut-off were fixed at 1 nm for the Lennard Jones parameters, and long range interactions were ensured by the Ewald summation with a precision of 10$^{-5}$.
\noindent MD simulations to determine diffusion coefficients from the mean square displacement and NEMD as well as Green-Kubo techniques were run in the ($N$,$V$,$T$) ensemble for 50 ns with a time step of 1 fs and each reproduced 5 times from different initial configurations. A Nose-Hoover thermostat were used to keep constant pressure and temperature during full simulations.\\ 
Binding energies presented in the article were determined from MD simulations in the ($N$,$V$,$T$) ensemble. Equilibrated structures from GCMC has been annealed during 10 ns, with a thermal ramp ranging from $T$=400K to 10K, with a time step of 1 fs. Binding energies were then determined from instantaneous energy calculations in annealed structures and have been reproduced 5 times to gain a better averaging in $E_b$.

\pagebreak

\begin{figure*}
\centering
\includegraphics[height=7cm]{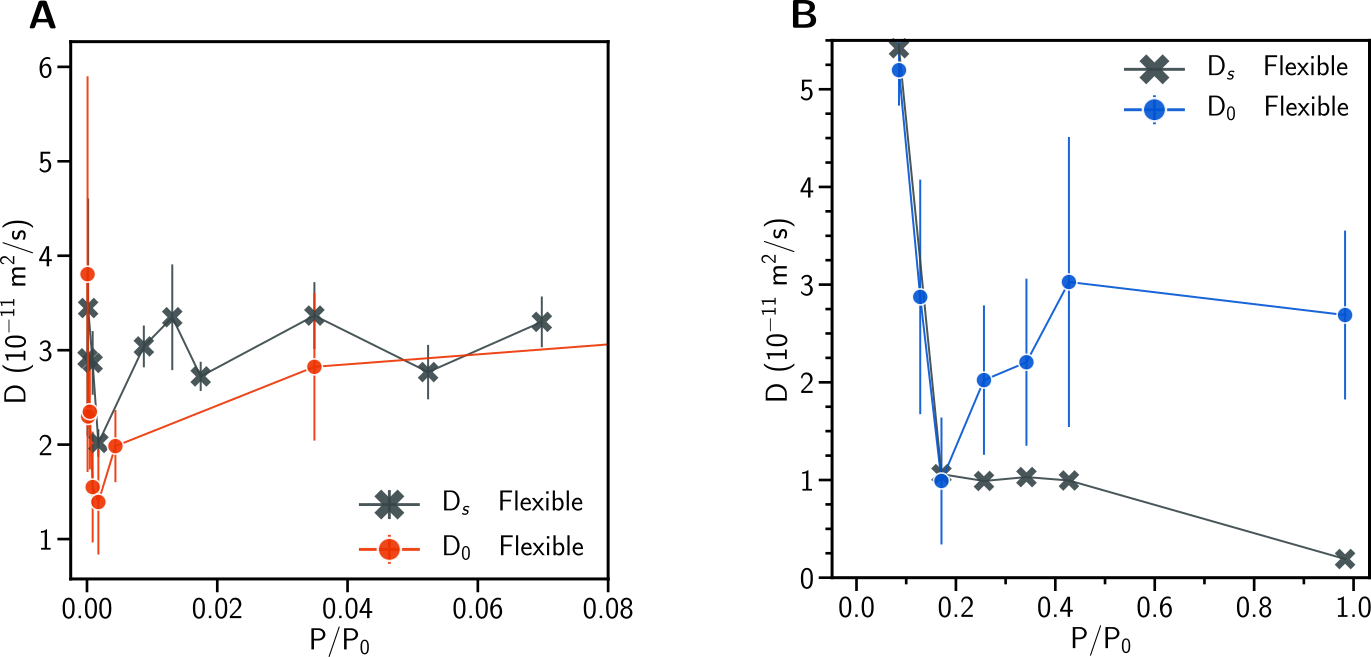}
\caption{A. CO$_2$ diffusion coefficients in flexible CALF-20. The dark crosses correspond to the self-diffusion in a structure model by the UFF force field, determined by the mean square displacement method. The red circles to the corrected diffusion with the UFF force field and the green squares to the corrected diffusion with the Dreiding force field. The latter has been determined from non equilibrium technique. B. Full red circles correspond to the corrected diffusion coefficient in flexible CALF-20 from the UFF force field. Open circles corresponds to the corrected diffusion coefficient in a rigid MOF structure.}
  \label{fig:figSI6}
\end{figure*}

\section*{Acknowledgments}
This work was supported by TotalEnergies S.E. through OneTech, Sustainability, CCUS R\&D program. On the behalf of all co-authors, Y. Magnin thanks T. Woo for sharing atomistic structure of CALF-20(Zn), B. Coasne for helpful discussions and TotalEnergies High Performance Computing Center for the CPU time provided on its super calculator PANGEA II. All authors also thanks TotalEnergies PERL for discussions concerning experiments.

\bibliographystyle{unsrt}  

\end{document}